\documentstyle[aps,prb,multicol,epsfig]{revtex}

\begin{document}
\input epsf
\draft
\title{Layer Analysis of the Structure of Water Confined in Vycor Glass}

\author{P.~Gallo, M.~A.~Ricci, and M.~Rovere\footnote[1]{Author to whom correspondence 
should be addressed; e-mail: rovere@fis.uniroma3.it}}
\address{Dipartimento di Fisica ``E.~Amaldi'', 
Universit\`a degli Studi ``Roma Tre'', \\ Istituto Nazionale per la Fisica della Materia,
 Unit\`a di Ricerca Roma Tre\\
Via della Vasca Navale 84, 00146 Roma, Italy.}
\maketitle

\begin{abstract}
A Molecular Dynamics simulation of the microscopic structure
of water confined in a silica pore is presented. 
A single cavity in the silica glass has been modeled as to reproduce the
main features of the pores of real Vycor glass. A layer analysis of the 
site-site radial distribution functions evidences the presence in the pore 
of two subsets of water molecules with different microscopic structure. 
Molecules which reside in the inner layer, close
to the center of the pore, have the same structure as bulk water 
but at a temperature of $30 K$ higher.
On the contrary the structure of the water molecules in the outer layer,
close to the substrate, is strongly influenced by the water-substrate 
hydrophilic interaction and sensible distortions
of the H-bond network and of the orientational correlations between 
neighboring molecules show up. Lowering the hydration
has little effect on the structure of water in the outer layer.
The consequences on experimental determinations of the structural properties
of water in confinement are discussed.

\end{abstract}


\date{\today}

\begin{multicols}{2}

\section{ Introduction}

The observation that in many real situations water is confined 
(in a rock, in a cell, in a microemulsion, in an ionic channel, 
in interstellar bodies etc.) 
explains the large interest devoted to structural and dynamical 
properties of water in 
restricted geometries. During the last decade indeed several experiments 
have been performed by using different techniques and different 
confining substrates, that have evidenced a rich 
variety of phenomena, sometimes even more puzzling than those 
known for bulk water.\cite{granada} Moreover recent theoretical studies have 
evidenced for confined water a thermodynamics more complex than that of 
bulk water.\cite{debene}  

In particular the comparison of the properties of bulk and confined water that 
emerges from different experiments is sometimes misleading. 
As a matter of fact 
the slowing down of the translational and rotational motion upon 
confinement, evidenced by nuclear magnetic resonance (NMR)\cite{denisov} and 
quasi elastic neutron scattering (QENS),\cite{mcbel1} apparently 
contrasts with 
the lower average density deduced for confined water with respect to the bulk 
state by small angle neutron scattering (SANS)\cite{benham}. Moreover neutron 
diffraction (ND) experiments on pure $D_2O$ have been interpreted in terms 
of an increased number of hydrogen bonds (HB) 
upon confinement;\cite{mcbel2} while 
experiments exploiting the isotopic H/D substitution on water 
reveal a reduction 
of such bonds and a lower tetrahedral order,\cite{vycor1,vycor2} in agreement 
with X-ray diffraction.\cite{mcbel2} On the 
other hand the analysis of the temperature derivatives of 
the radial distribution 
functions (RDF) of confined and bulk $D_2O$ suggests that the structural 
rearrangement upon lowering the temperature is the same for the 
two systems.\cite{dore}
More recently the idea has been put forward that lowering the amount of water 
inside the confining matrix is equivalent to lowering the temperature in bulk 
water.\cite{mcbel2,mcbel3,chen} Finally, besides and in partial disagreement 
with these interpretations of the experimental findings, two experimental 
evidences are well established: 
\begin{itemize}
\item{Confined water can be easily supercooled down to lower 
temperatures compared to the bulk.}
\item{The lower is the hydration level, the lower is 
the ice nucleation temperature.} 
\end{itemize}

The observation that the density profile of water in restricted geometries 
is not uniform,\cite{rossky,spohr,fang,robinson,jmol1,jmol2} whatever 
the substrate-water interaction is, 
suggests that most of the misfits between the experimental observations 
reported above are 
due to averaging the response of different water layers, 
that cannot be easily isolated 
in the experiments, unless an accurate study as a function of 
the hydration level is performed.
Molecular dynamics (MD) simulations performed in realistic environments 
can be then of great help in better interpreting the experimental data. 
As a matter of fact the system 
that has been more widely used by the 
experimentalists\cite{mcbel1,benham,vycor1,vycor2,mcbel3,chen,zanotti1,zanotti2,venturini} 
to confine water, namely porous Vycor glass,\cite{cornig} has also the advantage of being a simple molecular system, that can be computer 
simulated.\cite{jmol1,vellati} In the attempt of giving an 
harmonic interpretation of 
at least the structural properties of water in restricted geometries, we 
have performed a layer analysis of the microscopic structure of 
water in Vycor glass, at different 
temperatures and hydrations.

In Sec.~II we briefly describe the simulation method. In Sec.~III
we discuss how the calculation of the pair correlation functions
can be performed by taking into account the finite volume effects.
In Sec.~IV we introduce the layer analysis and present the main results.
The last section is devoted to the conclusions. 

\section{Molecular dynamics of confined water}

The behavior of water confined in the pores of Vycor glass 
is influenced by both the geometrical effects and the 
interaction with the hydrophilic surfaces. In order to account for these 
effects in computer simulation we constructed a single pore of
silica which reproduces
the most relevant average features of the pores of real Vycor glass. 
As described in more detail in a previous work~\cite{jmol1}, 
we carved a cylindrical pore of $40$~\AA\ diameter 
in a cubic cell of silica glass obtained by the usual procedure
of MD simulation.
Then we prepared the cavity surface by
removing the silicon atoms bonded 
to less than four oxygens. 
The oxygen atoms bonded to only
one silicon ({\it non bridging oxygens}) were saturated with
acidic hydrogens. 
We note that since the surface of the cavity is rough
its actual volume $V_p$ is
unknown and only roughly approximated to
a lower value by the volume $V_c$ of a cylinder of radius  $R_c=20$~\AA.

Inside the pore a fixed number $N_W$ of water molecules is inserted
and the MD simulation is performed at constant
density $\rho=N_W/V_c$. We assume, according to the 
experiments\cite{benham,vycor1}, that at full hydration 
the density is $\rho_f=0.0297$~\AA$^{-3}$, corresponding to $N_W=2600$.
In the following we will present results for $N_W=2600$ and $N_W=1500$
which is roughly the half hydration case. 
 
Water-water and water-substrate interactions are respectively described 
by the SPC/E site model\cite{spce} and by an empirical 
potential model, consisting of Lennard-Jones and Coulomb forces.
The potential parameters are 
those used and tabulated in Ref~\onlinecite{jmol1} where further
technical details can be found.
During the MD runs the atoms of the Vycor glass are kept fixed.  
Periodic boundary conditions are applied along the axis of the cylinder ($z$-
direction). The motion is confined in the $x$-$y$ plane. 

\begin{figure}
\centering\epsfig{file=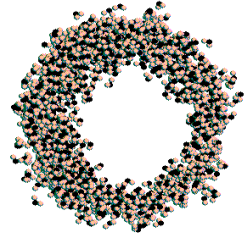,width=0.75\linewidth}
\caption{ Projection on the x-y plane of a 
snapshot from simulation of $1500$ molecules of
pure SPC/E water inside a cylindrical
cavity of Vycor glass. The radius of the cavity is $20$~\AA. For
the sake of clarity only water molecules are shown. Oxygens are black and
hydrogens are gray.
\protect\label{fig:fig0}}
\end{figure}

A snapshot from the simulation of $1500$ water molecules is presented in 
Fig.~\ref{fig:fig0}. Only water molecules are displayed, the hydrophilic
nature of the Vycor surface is evident, since all water molecules
are attracted toward the substrate.

%______________________________________________________________________________
\section{Excluded volume effects on the pair correlation functions}

The calculation of the site-site radial distribution functions (SSRDF) 
in computer simulation is not 
straightforward when dealing
with a confined system, since excluded volume effects must be
carefully taken into account.~\cite{vycor2,jmol2,soper,branka}  

From the MD simulation we can
calculate in the usual way the average number 
$n_{\alpha \beta}^{(2)}(r)$ of
sites of type $\beta$ lying in a spherical shell $\Delta v(r)$ at
distance $r$ from a site of type $\alpha$. Then
normalizing to the average number of atoms of an ideal gas at
the same density in the same spherical shell, one obtains~\cite{allen}
\begin{equation}
\tilde{g}_{\alpha \beta} \left( r \right) = 
\frac{ n_{\alpha \beta}^{(2)}(r) }{ {\frac{N_\beta }{V_p}} 
\Delta v(r)} \ ,
\label{grmd}
\end{equation}
where $N_\beta$ is the total number of $\beta$ sites and $V_p$ is the
volume of the simulation cell. 

At variance with a bulk
liquid, if we consider a collection of
non-interacting particles in a confining volume, where periodic
boundary conditions are, for instance, only along the $z$-axis, as
in our case, the
\emph{uniform} radial distribution function is not 
simply $g_u(r)=1$. It indeed depends on the geometry of the
confining system and generally will not be a constant.
Hence the functions defined in Eq. ~(\ref{grmd})  
must be further normalized to the uniform profile $g_u(r)$.
In our case this is given by~\cite{jmol2}: 
\begin{equation}
g_u(r)=\frac{V_p}{\left( 2\pi \right)^3} \int d^3 Q P_{cyl} \left(Q \right)
\label{gru}
\end{equation}
$P_{cyl}(Q)$ is the form factor of the cylindrical simulation
cell of height $L$ and radius $R_c$~\cite{glatter} 
\begin{equation}
P_{cyl} \left( Q \right) =
\int_0^1 d\mu \left[ j_0 \left( \frac{\mu Q L}{2} \right) \right]^2
\left[ \frac{2 j_1 \left( Q R_c \sqrt{1-\mu^2} \right)}{Q R_c\sqrt{1-\mu^2}}
\right]^2 
\label{pcyl}
\end{equation}
where $j_n(x)$ are the Bessel's functions of order $n$. 

The properly normalized SSRDF are then obtained as:
\begin{equation}
g_{\alpha\beta}(r)=\frac{
\tilde{g}_{\alpha\beta}(r)}  {f_c \cdot g_u(r)}.
\label{grcorr}
\end{equation}
The correction factor $f_c$
accounts for the roughness of the surface and 
the already discussed uncertainty in the volume determination. It is
adjusted to give the corrected pair correlation functions
oscillating around 1 at large $r$.

\begin{figure}
\centering\epsfig{file=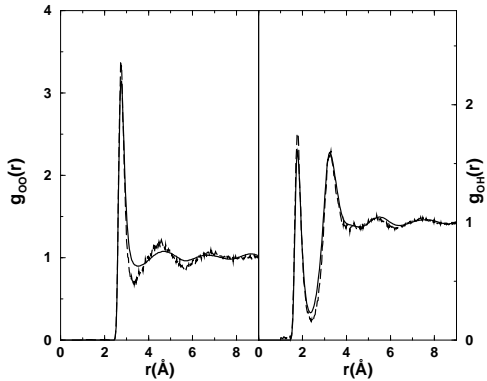,width=0.75\linewidth}
\caption{Site-site radial distribution functions $g_{OO}(r)$
(on the left) and $g_{OH}(r)$ (on the right) at full hydration
at $T=298 K$ (full line) compared with the same functions of the
bulk water (long dashed line). The functions of the confined system
have been calculated by using Eqs.~(\ref{grmd})-(\ref{grcorr}). 
\protect\label{fig:grtot}}
\end{figure}

The corrected SSRDF for the full hydration
case are shown in
Fig.~\ref{fig:grtot} and are compared to the corresponding functions 
of SPC/E water at ambient
conditions. As discussed in a previous work~\cite{jmol2}
the modifications of the oxygen-oxygen (OO) and of the oxygen-hydrogen (OH)
functions relative
to bulk water are in good qualitative agreement with the 
experiments.~\cite{vycor2}
For instance the first minimum of the OO function becomes
shallower and fills in and the H-bond peak becomes less intense
upon confinement.  

%%%%%%%%%%%%%%%%%%%%%%%%%%%%%%%%%%%%%%%%%%%%%%%%%%%%%%%%%%%%%%%%
%_______________________________________________________________

\section{Layering effect and structure of confined water}

Looking at the density profile reported in the inset of 
Fig.~\ref{fig:comp015} as a function of $R=\sqrt{x^2+y^2}$ we see that
the density profile is very flat in the range $0 <R< 15$~\AA\
and its value is in agreement with the density of confined
water at full hydration observed in the experiments.
For $R>15$~\AA\ water adsorbs and a
double layer structure is formed close to the surface.
The double layer with density higher than the average extends 
for almost $3$~\AA. 
A depletion layer of $2$~\AA, due to the short range repulsion
of the substrate is also visible.

We notice that the density profile is almost independent of the
temperature.

As a consequence of the hydrophilic interaction a strong distortion
of the hydrogen bond network of water in the layers close to
the substrate is observed.~\cite{jmol1,jmol2}
Moreover this layering effect has important consequences on the
dynamical behaviour of water as recently shown.~\cite{gallo-prl}
It seems therefore appropriate to develop a layer analysis for the
SSRDF by separating the contribution of the water molecules
which spend most of the time in the internal layer ($0 <R< 15$~\AA )
from the contribution of the molecules belonging to the double layer
or to the depletion layer. 

\subsection{Inner layer.}

The SSRDF in the inner layer can be obtained from the MD data by 
using Eqs.(~\ref{grmd})-(\ref{grcorr}) where
now the radius of the cylinder is $R_c=15$~\AA.  

\begin{figure}
\centering\epsfig{file=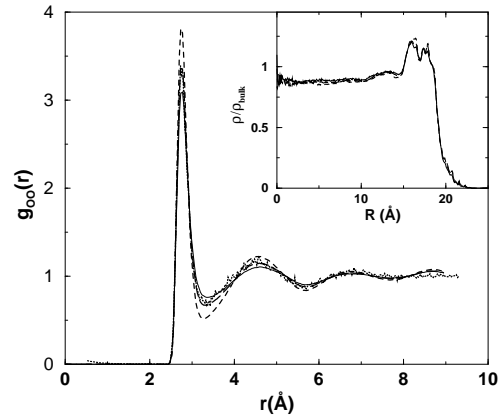,width=0.75\linewidth}
\caption{Site-site radial distribution functions $g_{OO}(r)$
at full hydration
calculated for the inner layer ($0 <R< 15 $~\AA, see text) 
at temperatures $T=298 K$ (continous line), $T=270 K$ (long dashed
line) and $T=240 K$ (short dashed) compared
with the corresponding function of the bulk at ambient
conditions (dotted line). The SSRDF of bulk water at ambient
conditions is almost coincident with that of confined water at $270 K$.
In the inset the density profile at
full hydration for $T=298 K$, $T=270 K$ and $T=240 K$ are
reported by using the same symbols.
\protect\label{fig:comp015}}
\end{figure}

In Fig.~\ref{fig:comp015} we show the
resulting $g_{OO}(r)$ for the inner layer at different temperatures
and compare with the equivalent function of ambient temperature
SPC/E water.

We notice that
the $g_{OO}(r)$ of the confined system are very similar to those
of bulk water. In particular present data at $T=270$~K superimpose
to the oxygen-oxygen RDF calculated for bulk water at ambient
conditions. For the other two SSRDF (not shown) we find the same
result, suggesting that as far as the microscopic structure is concerned
water confined in the middle of the pore behaves as bulk water at an 
higher temperature ($\Delta T\sim 30 K$). This result may explain why
confined water can be more easily supercooled than the bulk.  

\subsection{Outer layer.}

In the calculation of the SSRDF for the outer layer 
the geometry of the confining volume and the form factor 
appearing in Eq.(\ref{gru}) are different from those used in the previous case.
Now the function $g_u(r)$ must be calculated in the region
$15 <R< 20$~\AA\ where the water is adsorbed on the
surface. In doing so, we can assume that the uniform density of
non interacting atoms is different from zero and
constant in the range $15<R<18$~\AA, neglecting the
small contribution which comes from particles in the depletion 
layer $18 <R< 20$~\AA.
The form factor to be used
is now that of two concentric cylinders of radius
$R_1=15$~\AA\ and $R_2=18$~\AA\ respectively:
\begin{equation}
P_{diff}(Q)= \frac{1}{\left( V_2 - V_1 \right) } \left[
V_2 P^{(1)}_{cyl}(Q) - V_1  P^{(2)}_{cyl}(Q) \right]
\label{pdiff}
\end{equation} 
where $V_i$ is the volume of the cylinder of radius $R_i$ and
$P^{(i)}_{cyl}(Q)$ is given by Eq.(~\ref{pcyl}) with $R_c=R_i$.

\begin{figure}
\centering\epsfig{file=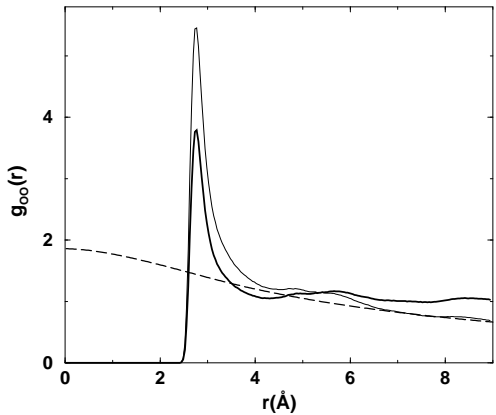,width=0.75\linewidth}
\caption{Site-site radial distribution functions $g_{OO}(r)$
at full hydration
calculated for the outer layer ($15 <R< 20 $~\AA ) 
at temperatures $T=298 K$.
Thinner solid line is $\tilde{g}_{OO} (r)$ defined in Eq.~(\ref{grmd});
Dashed line is the uniform profile calculated according to Eq.~(\ref{gru}).
The corrected RDF is reported as a thick solid line.  
\protect\label{fig:gr1520OO}}
\end{figure} 

\begin{figure}
\centering\epsfig{file=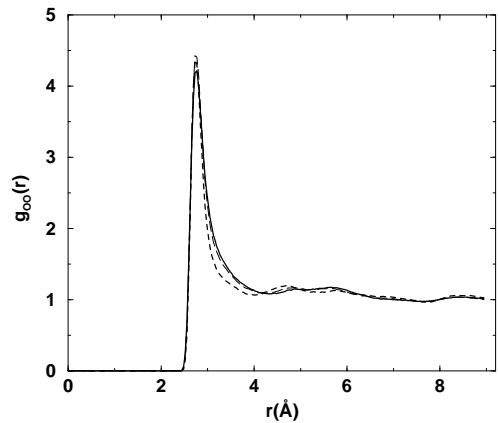,width=0.75\linewidth}
\caption{Site-site radial distribution functions $g_{OO}(r)$
at full hydration
calculated for the outer layer ($15 <R< 20 $~\AA ) 
at temperatures $T=298 K$ (full line), $T=270 K$ (long dashed
line), $T=240 K$ (short dashed line).
\protect\label{fig:gr1520OOT}}
\end{figure}

In Fig.~\ref{fig:gr1520OO} we report the RDF of the oxygen sites  
obtained with 
this procedure together with the initial distribution $\tilde{g}_{OO} (r)$
calculated according to Eq.~(\ref{grmd}) and the 
uniform profile $g_u(r)$ for the concentric cylinders.
We notice that the correction for the excluded volume effects may still be
not completely satisfactory in the region of the main peak,
due to neglecting the presence
of the depletion layer. In spite of this we
can infer that the structure of water in the double layer
close to the substrate is strongly distorted with respect to the bulk. 
The $g_{OO}(r)$
function does not show any feature typical of the tetrahedral
arrangement of H-bonded molecules. In particular the
second peak is shifted to higher distances and its intensity is 
much reduced with respect to the bulk water case. 
We observe also that the $g_{OO}(r)$ of the outer layer does not
change much with the temperature, as can be seen in Fig.~\ref{fig:gr1520OOT}:
only a sharpening of the main peak at the lowest temperature 
is visible.

\begin{figure}
\centering\epsfig{file=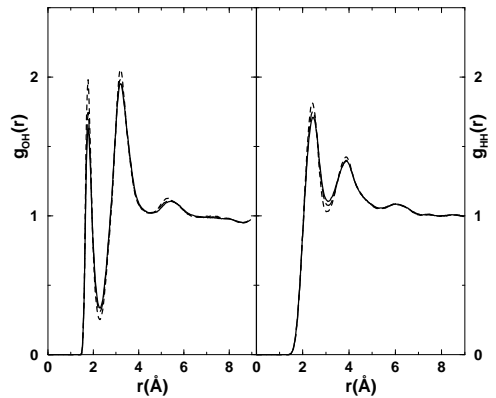,width=0.75\linewidth}
\caption{Site-site radial distribution functions $g_{OH}(r)$
(on the left) and $g_{HH}(r)$ (on the right) 
at full hydration
calculated for the outer layer ($15 <R< 20 $~\AA ) 
at temperatures $T=298 K$ (full line), $T=270 K$ (long dashed
line), $T=240 K$ (short dashed line).
\protect\label{fig:gr1520OHT}}
\end{figure}

The $g_{OH}(r)$ and $g_{HH}(r)$ functions calculated for the 
outer layer are reported in Fig.~\ref{fig:gr1520OHT}
at different temperatures. At variance with the $g_{OO}(r)$
these functions show the main peaks at about the same $r$-values 
as their analogous functions obtained for bulk water. Yet  
differences are observed in the relative intensity of the first two 
peaks: these may be partially ascribed to residual excluded 
volume effects due 
to the presence of the depletion layer.
Also the $g_{OO}(r)$ and $g_{OH}(r)$ functions are weakly 
temperature dependent.

\subsection{Half hydration.}

The density profile calculated at half hydration ($N_W=1500$) evidence
the presence of a double layer of adsorbed water molecules, with 
the same characteristics as in the full hydration case. Thus
we can calculate the SSRDF for the outer layer by using the 
uniform profile calculated in the previous section for the full 
hydration case (Eq.\ref{pdiff}).

\begin{figure}
\centering\epsfig{file=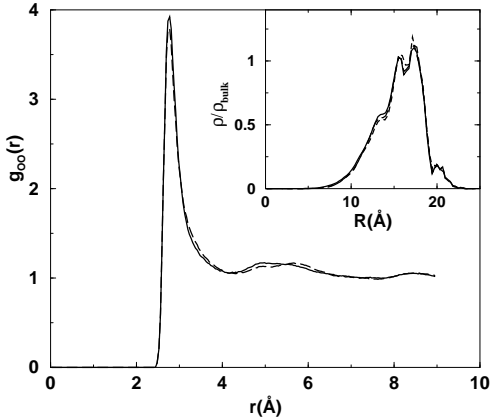,width=0.75\linewidth}
\caption{Site-site radial distribution functions $g_{OO}(r)$
calculated for the outer layer ($15 <R< 20 $~\AA )
at half hydration (full line) compared with the
same function at full hydration (long dashed line) 
at temperatures $T=298 K$.
In the inset the density profile at
half hydration for $T=298 K$ (full line), $T=270 K$
(long dashed line) and $T=240 K$ (short dashed line) are
reported. 
\protect\label{fig:gr1500-OO}}
\end{figure} 

\begin{figure}
\centering\epsfig{file=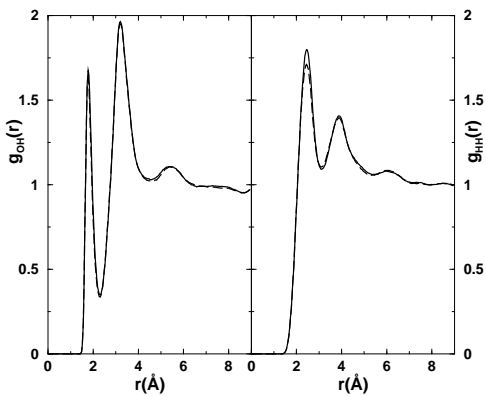,width=0.75\linewidth}
\caption{Site-site radial distribution functions $g_{OH}(r)$
(on the left) and $g_{HH}(r)$ (on the right)  
at half hydration (full line) 
calculated for the outer layer ($15 <R< 20 $~\AA )
compared with the same functions at full hydration (long dashed line) 
at temperature $T=298 K$ (full line).
\protect\label{fig:gr1500-OH-HH}}
\end{figure}

In Fig.~\ref{fig:gr1500-OO} we compare the $g_{OO}(r)$ functions 
at the two hydration levels investigated, at room temperature. 
In Fig.~\ref{fig:gr1500-OH-HH} the same comparison is done for the $g_{OH}(r)$
and $g_{HH}(r)$ functions. 
It is clear from these figures that changing the hydration does not
affect the SSRDF of the outer layer; the same evidence is found 
at the other investigated temperatures.

The SSRDF of the inner layer at half hydration are not presented, since the
finite volume corrections cannot be calculated with enough accuracy 
when the density drops too rapidly (see the inset of 
Fig.~\ref{fig:gr1500-OO}) 

\section{Summary and Conclusions}

We have presented the results of a layer analysis of the
pair correlation functions of confined water obtained by computer 
simulation. Water is strongly adsorbed, 
due to the hydrophilic interaction with the substrate.
Its density profile, which is almost independent of 
the temperature, shows the presence of a double layer 
of approximately $5$~\AA\ close to the substrate that we call
the outer layer. 
At full hydration water fills the pore and we can 
calculate the structure of the set of molecules in the
inner layer by excluding the molecules in the outer layer.
The SSRDF of the molecules in the inner layer are very
similar to those of bulk water. In particular the
structure of the confined water in the inner layer 
at $T=270 K$ almost
coincides with the structure of bulk water at ambient temperature, thus
explaining why confined water can be supercooled down to lower 
temperatures than bulk water.

On the contrary the microscopic structure of water belonging to the outer 
layer differs from that of 
bulk water. In particular the oxygen-oxygen pair correlation function
does not show any signature of the tetrahedral arrangement
characteristic of bulk water. This is in agreement with 
experimental findings~\cite{vycor2} and supported by
previous computer simulation analysis of the distortion of the
hydrogen bond network of confined water.~\cite{jmol2}
Moreover the SSRDF calculated for the outer layer are 
almost insensitive to temperature and hydration level. The 
first observation explains 
why the temperature derivatives of the RDF look so similar to those 
of bulk water\cite{dore}. 
The second suggests that the structure of water in the inner 
layer can be safely studied in a real diffraction experiment, 
looking at the difference between the data collected at different 
hydration levels. We stress finally that  
the analysis of the dynamical 
behavior of the molecules confined in the outer layer suggests that they 
are in a glassy state already at ambient temperature.~\cite{gallo-prl}
As a consequence their relative orientations must be  strongly distorted 
relative to the bulk, in agreement with what found in the present analysis.
The absence of tetrahedral order in the outer layer is also responsible 
for the lower nucleation temperature upon lowering the hydration level.

%\section {Acknowledgements}

\end{multicols}

\end{document}